\documentclass[prb,aps,twocolumn,floatfix,amsmath,amssymb,superscriptaddress,tightenlines]{revtex4}
\usepackage{graphicx}
\usepackage{epstopdf}
\newcommand{\be}{\begin{equation}}
\newcommand{\ee}{\end{equation}}
\usepackage{amsfonts}
\usepackage{bm}
\usepackage{color}
\begin{document}

\date{\today}
\title{Anomalies in the Entanglement Properties of the Square Lattice Heisenberg Model}
\author{Ann B. Kallin}
\affiliation{Department of Physics and Astronomy, University of Waterloo, Ontario, N2L 3G1, Canada}

\author{Matthew B. Hastings}
\affiliation{Duke University, Department of Physics, Durham, NC, 27708}
\affiliation{Microsoft Research, Station Q, CNSI Building, University of California, Santa Barbara, CA, 93106}

\author{Roger G. Melko}
\affiliation{Department of Physics and Astronomy, University of Waterloo, Ontario, N2L 3G1, Canada}

\author{Rajiv R. P. Singh}
\affiliation{Physics Department, University of California, Davis, CA, 95616}

\begin{abstract}
We compute the bipartite entanglement properties of the spin-half square-lattice Heisenberg model by a variety of numerical techniques that include valence-bond quantum Monte Carlo (QMC), stochastic series expansion QMC, high temperature series expansions and zero temperature coupling constant expansions around the Ising limit. We find that the area law is always satisfied, but in addition to 
the entanglement entropy per unit boundary length, there are other
terms that depend logarithmically on the subregion size, arising from broken 
symmetry in the bulk and from the existence of corners at the boundary. We find that the numerical results are anomalous in several ways. First, the bulk term arising from broken symmetry deviates from an exact calculation that can be done for a mean-field N\'eel state. Second, the corner logs do not agree with the known results for non-interacting Boson modes. And, third, even the finite temperature mutual information shows an anomalous behavior as $T$ goes to zero, suggesting that $T \rightarrow 0$ and $L \rightarrow \infty$ limits do not commute.
These calculations show that entanglement entropy demonstrates a very rich behavior
in $d>1$, which deserves further attention.
\end{abstract}

\maketitle

\section{Introduction}

The study of bipartite entanglement properties in quantum statistical models
is a promising way to understand and classify their topological and universal 
properties.\cite{Cardy,ALreview} In one spatial dimension, the existence of finite entanglement
entropy for gapped systems and a logarithmically divergent entanglement 
entropy for gapless systems is well understood. The coefficient of the log
divergence is universal in that it depends on the central charge and not
on the microscopic details of the system.
In two dimensions (2D), several gapless systems have been shown to display an 
area law, where the entropy per unit boundary length is necessarily non-universal,
reflecting all the microscopic degrees of freedom at the boundary.
Thus, looking for universal behavior necessitates looking for subleading
terms. The area law coefficient will have a subleading power law dependence
on the size of the system. However, more interestingly, there can be terms
associated with corners on the boundary, with broken symmetry, 
or with non-trivial topology in the bulk which could be universal and allow
one to classify different topological and critical phases.

The spin-half square-lattice Heisenberg model with nearest-neighbor 
interactions is one of the most studied and best understood models in 2D.\cite{CHN} 
The model has long-range order, and spin rotational symmetry
is spontaneously broken in the ground state. This broken symmetry is
well described by a tower of rotor states,\cite{tower} whose energy scales
with system size $L$ as $1/L^d$ in $d$ dimensions. 
For large systems, these states
are well separated from excitations around the ground state,
which in the long-wavelength limit are 
spin-waves whose energies scale as $1/L$. These spin-waves are
believed to become non-interacting in the long-wavelength limit
as long as one is away from a quantum critical point where the  
long-range order might go continuously to zero.

In this paper, we have developed a number of different computational methods to
calculate entanglement properties, which are valid for arbitrary
dimensional quantum statistical models. Stochastic series
expansion (SSE) quantum Monte Carlo (QMC)\cite{SSE,SSE1} and High Temperature Expansions (HTE) are
methods that allow one to calculate thermodynamic properties
of the model at finite temperatures, and can be used to
obtain the Renyi mutual information associated with dividing the
system into two regions $A$ and $B$.\cite{XXZ} This mutual information
should reduce to twice the entanglement entropy as $T\to 0$.
The Valence Bond (VB) QMC 
method is a particularly powerful tool
for studying properties of the model directly at $T=0$.\cite{Sandvik,Beach,Beach2} 
We have developed extensions of this method which employ loop updates\cite{AWSloop} 
necessary to accurately
calculate the entanglement properties for finite lattices
with different bipartite divisions. Finally, Ising expansions at
$T=0$ provide yet another method to calculate properties
of the Heisenberg model in its ground state. By approaching the ground
state of the Heisenberg model from the ordered side one can
calculate a specific ordered state and its entanglement properties.

Our first result is an estimation of the leading non-universal area law coefficient,
calculated as {$a=0.0965\pm0.0007$} in VB QMC (Section~\ref{VBQMCsec}) and $a= 0.094 \pm 0.001$ in 
Ising series expansions (Section~\ref{HTEsec}), 
which are in
good agreement. The small discrepancy shows that there are systematic
errors not included in the numerical estimates of error bars. However, these are
of order one percent.

The universal contributions to the subleading scaling come from a number of sources.
For example, one might
expect the universal entanglement properties of the model to be
related to the broken symmetry of the N\'eel state, and to the presence of free Bosons
resulting from non-interacting spin-waves.
The entanglement properties of free Bosons (and free Fermions) have been
computed in 2D.\cite{EEstat}  Furthermore, the entanglement
properties of a mean-field N\'eel state, where all spins on sublattice 1
and sublattice 2 separately form a maximal spin state that are then
combined into a singlet, can be calculated exactly. 
This gives a mean-field bulk entanglement entropy associated
with broken symmetry which scales as $c\ln(\ell)$, with $c=2$, where $\ell$ is the length of the boundary (Section \ref{sec:mft}).  In contrast, our
VB QMC simulations for a bipartite division with no corners
show a logarithmic term with {$c= 0.74\pm 0.02$}. Recently, the entanglement
properties of this model were calculated using modified spin-wave theory,\cite{yalegroup}
which gave an estimate of $c=0.92$.  Thus, our numerical simulations are
much closer to spin-wave theory.  This suggests that, in addition to
broken symmetry, there are logarithmic contributions in the bulk
that come from other sources.

The corner contributions can be obtained in QMC by comparing the log terms in a system using a  boundary with corners to a system without boundary corners, giving an estimate of {$-0.10\pm 0.02$}. 
The series estimate, $-0.080 \pm 0.008$,
gives reasonable agreement with the QMC results.
In contrast, if one takes two free Boson modes contributing to the corners, then one gets $\sim -0.0496$ (Section~\ref{SWTsec}).

Finally, we have studied the properties of the model at
finite temperatures, and examined the approach to $T\to 0$. This
can be done by SSE QMC and HTE. The HTE extrapolations for the Renyi mutual information agree well
with QMC  down to $T\approx1$. Below this temperature, the QMC data
shows a sudden decrease and a crossover to a lower saturation value
as $T\to 0$. The latter is consistent with the entanglement entropy
calculated at $T=0$. The HTE does not show any sharp decrease. The
sharp decrease has a size dependence and could imply the limits
of $T\to 0$ and $L\to \infty$ do not commute. Such non-commutation
is well known for other response functions of the Heisenberg 
model\cite{DSFisher,Hasenfratz}
and also suggests a sizeable non-mean-field contribution to the area
law term.

\section{Valence Bond Quantum Monte Carlo Algorithms and Results}  \label{VBQMCsec}

In order to calculate the zero-temperature scaling of the Renyi entanglement entropy in the Heisenberg model on finite-size lattices, we employ the valence-bond (VB)
quantum Monte Carlo (QMC) method developed by Sandvik.\cite{Sandvik}  This is a highly-efficient method to project out the model's ground state by repeated application of the Hamiltonian to a trial wavefunction, through a Monte Carlo sampling
of bond operators.
As an improvement upon our previous entanglement measurement procedure,\cite{swap} we use a modified version of the more efficient loop algorithm.\cite{AWSloop} This modification allows for a change to occur in the Monte Carlo weight, by modifying the space-time topology of the 
simulation cell.
As in Ref.~[\onlinecite{swap}], this modified weight is required so that one can measure the {\it difference} between entanglement entropies of two distinct system subdivisions, instead of the absolute entanglement entropy.  
Then, if the geometry of the regions are chosen properly the difference can converge faster than the bare entanglement entropy measurement.

\subsection{Valence Bond Quantum Monte Carlo}

First we briefly discuss the basic VB QMC algorithm (for more detail see Refs.~[\onlinecite{Sandvik,Beach,Beach2}]) and the more recently developed loop update (see Ref.~[\onlinecite{AWSloop}]) which significantly improves the scaling of the algorithm.
The foundation of the VB QMC technique is to project out the ground state of the system, done by applying a high power of the Hamiltonian $\mathcal{H}^M$ to a trial state.
We use the Heisenberg Hamiltonian, rewritten in terms of bond operators $(H_{ab} = \tfrac{1}{4} - {\bf S}_a \cdot {\bf S}_b)$ acting on pairs of sites ($a$ and $b$), which are nearest-neighbor pairs in this paper.
The Hamiltonian to the power $M$ can be written as a sum of possible arrangements of these bond operators in a list of size $M$.
The Monte Carlo algorithm importance samples terms in this sum, using a weight which depends on the number of off-diagonal operators in the term.\cite{Sandvik}

In its original form, the VB QMC scheme can be used to project out one copy of the ground state wavefunction (single-projector) or simultaneously project two copies (double-projector) which can be used to measure expectation values of observables in the simulation.\cite{Sandvik,Beach,Beach2}
Our previous scheme for measuring Renyi entanglement entropy\cite{swap} employed a double-projector method for calculating the
expectation value of a {\it Swap} operator.  In this paper, we develop a highly-efficient loop variation of this measurement algorithm, outlined below.

\subsubsection{The Loop Algorithm\cite{AWSloop}}

The loop update for VB QMC simulations was introduced in Ref.~[\onlinecite{AWSloop}] as a highly efficient way of carrying out the 
sampling procedure.  In addition to working in a basis of valence bonds, this scheme also samples over spin states.  This combined spin/bond basis is shown to eliminate the need for a rejection step, and thus samples operators and basis states with high efficiency. 

To begin, operators in this case are divided into two classes,
\begin{eqnarray}
	H_{ab}(1) &=&(\tfrac{1}{4} - S^z_aS^z_b) \\
	H_{ab}(2) &=&     -\tfrac{1}{2}(S_a^+S_b^- + S_a^-S_b^+)
\end{eqnarray}
called diagonal and off-diagonal operators respectively, where the sum $H_{ab}(1) + H_{ab}(2)$ is equal to the bond operators $H_{ab}$ from the standard VB QMC algorithms mentioned above.\cite{Sandvik,Beach,Beach2}

\begin{figure} {
\includegraphics[width=3.0 in]{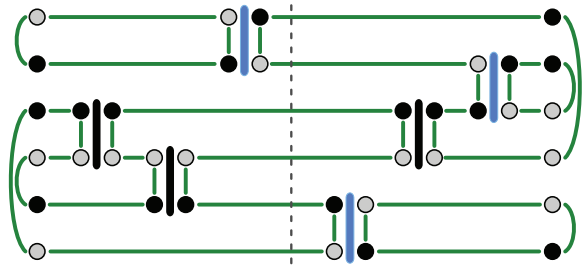}
\caption{A possible simulation cell diagram for a 6-site system, including loops, operators, the initial valence bond states, and the compatible initial spin states.
Up (down) spins are shown in grey (black). (Off-) diagonal operators are shown in (blue) black, giving a total of $M=3$ operators in this example.
The dashed line denotes $\lvert V_L \rangle$ (propagated from the trial state on left)  and $\lvert V_R \rangle$ (similarly from the right).
\label{loop1} }
} \end{figure}

The loop algorithm is best visualized using a diagram of the simulation cell showing the placement of valence bonds, spins, and operators, as depicted in Figure \ref{loop1}.
This diagram represents two VB trial states, which are the left and right edges of the figure, each projected ``inwards'' (by $M=3$ operators).  The projected state occurs in the center of the diagram, denoted by the dashed line.
Along with the trial VB states, initial spin states are selected at random, with the condition that the two spins in a single VB must be antiparallel.
For each trial state, $M$ operators are chosen such that they  each act on a pair of antiparallel spins (in the initial step of the algorithm, these are all diagonal operators).
There are then two types of updates in this algorithm: spin updates and operator updates.

For the spin update, loops are first constructed by linking the operators and valence bonds (shown in Fig.~\ref{loop1}).
Then, for each loop that is built, a decision is made to flip all the spins in that loop, with probability 1/2.
This update samples possible spin states for the given valence bond configuration.
In the second type of update, the operators in the list are changed so that diagonal operators are re-sampled at random, subject to the condition that they remain acting on antiparallel sites.
This reconfigures the propagated valence bond states and the topology of the simulation cell loops for future updates.

Measurements can be computed as usual\cite{Sandvik} using the propagated valence bond states, $\lvert V_L \rangle$ and $\lvert V_R \rangle$, which can be extracted from the simulation cell diagram by following the loops crossing the dotted line in Fig.~\ref{loop1}.  

In the next section we discuss the measurement of Renyi entanglement entropies with either the double-projector or loop algorithms.
In the following section we describe this measurement with the loop algorithm using a modified VB QMC simulation cell.

\subsection{Measuring Renyi Entropies with VB QMC}
Numerical techniques such as exact diagonalization or density matrix renormalization group (DMRG) simulations are able to directly measure entanglement entropy, since the calculation provides access to the density matrix.  In contrast, it is typically a challenge to measure entanglement entropy with QMC, as the density matrix is not sampled in a straightforward way.  
Over the past few years there have been several proposals of entanglement measures in Monte Carlo simulations.\cite{Alet, entanglement-fluctuations, loopentropy}
Recently, methods to measure Renyi entanglement entropy using a swap (or more generally a permutation)\cite{Quench} operator have been developed and implemented in several types of Monte Carlo methods.\cite{swap, XXZ} Next, we outline the basic methodology  relevant for VB basis QMC,\cite{swap} based on the expectation value of the {\it Swap} operator, before detailing current advances for improving the measurement efficiency using a hybrid loop-ratio trick estimator.

\subsubsection{Renyi  Entropies and the Swap Operator}

We are interested in the generalized Renyi entropies, which quantify entanglement between a system subdivided into two regions, $A$ and $B$.
They are defined as 
\begin{equation} \label{renyi}
S_{\alpha}(A) = \frac{1}{1-\alpha}\ln \left[ {\rm Tr}(\rho_A^{\alpha}) \right],
\end{equation}
where $\rho_A = {\rm Tr}_B (\rho)$ is the reduced density matrix of the total system traced out over region $B$ and the standard von Neumann entanglement entropy is recovered in the limit as $\alpha \rightarrow 1$.\\

Despite the inaccessibility of the full wavefunction of the system in Monte Carlo techniques, it is possible to sample ${\rm Tr}(\rho_A^{\alpha})$ for integer $\alpha > 1$.
This is accomplished by taking the expectation value of a $Swap$ operator\cite{swap} for $\alpha = 2$
or permutation operator for $\alpha > 2$;\cite{Quench} e.g.
\begin{eqnarray} 
S_2(A) &=& - \ln(\langle Swap_A \rangle)\\
S_3(A) &=& - \frac{1}{2}\ln(\langle \Pi_3^A \rangle).
\end{eqnarray} 

To measure the $\alpha^{\rm th}$ entropy, each projected state must be composed of $\alpha$ non-interacting copies of the system.
The permutation operators $\Pi_\alpha^A$ act to cyclicly exchange the state in region $A$ between the $\alpha$ different copies of the system, and are constructed such that  $\langle \Pi_\alpha^A \rangle = {\rm Tr}(\rho_A^{\alpha})$.
In the case of spin states (which for each MC step in the simulation is a product state) we can simply swap the states within region $A$ between copies of the system.  
For valence bond states the application of the permutation operator also has a simple result; it acts to exchange the endpoints of valence bonds within region $A$ between copies of the system, and as such can create bonds between the non-interacting copies.\cite{swap}

The bare measurement of the $Swap$ operator has been shown\cite{swap} to have problems with convergence for large region $A$, while in principle the measurement should be symmetric such that $\langle Swap_A \rangle = \langle Swap_B \rangle$, since the two density matrices $\rho_A$ and $\rho_B$ have the same eigenvalues.
This is because the exchange of a larger region gives a larger number of different states as a result of that swap, and thus a larger range and number of possible values in the expectation value.
It simply takes more Monte Carlo steps to converge upon the same result.
Another way to think of it is that, even though ${\rm Tr}(\rho_A^{2}) = {\rm Tr}(\rho_B^{2})$, if region $A$ is much larger than region $B$ then ${\rm Tr}(\rho_A^{2})$ contains exponentially more terms than ${\rm Tr}(\rho_B^{2})$.
Getting the value to converge by a stochastic sampling of these terms takes much longer.  However, it is possible to instead sample a series of smaller regions and greatly improve the convergence time.  This technique is described in the following section.

\subsubsection{The Ratio Trick}
The convergence issue mentioned above can be addressed by a reweighting of the Monte Carlo sampling scheme.
Begin by considering the double-projector method, where one measures the expectation value of {\it Swap} by sampling terms from 
\be
\langle Swap_A \rangle =\frac {\sum_{lr} w_r w_r\langle V_l  \lvert V_r\rangle\frac{  \langle V_l  \lvert Swap_A \lvert V_r\rangle}{ \langle V_l  \lvert V_r\rangle}}
		{\sum_{lr} w_l w_r \langle V_l  \lvert V_r\rangle}
		\label{weight}
\ee
where $\lvert V_l \rangle, \lvert V_r \rangle$ are the states obtained by applying lists of bond operators to the trial states and $w_l , w_r $ are the weights accrued by applying those operators.
Terms are sampled proportional to the total weight $W = w_l w_r \langle V_l  \lvert V_r\rangle$, by accepting a new configuration with probability $W^{\rm new}/W^{\rm old}$.
Thus one can simply measure $\tfrac{  \langle V_l  \lvert  Swap_A \lvert V_r\rangle}{ \langle V_l  \lvert V_r\rangle}$ once per Monte Carlo step, and the average value will give us $\langle Swap_A \rangle$.

The convergence difficulties can be combatted by using the {\it ratio trick}.\cite{swap} One modifies the sampling weight to include the expectation value of a $Swap$ operator for a region $A$ that is close in size to the region we intend to measure.  
One can then measure the ratio of these two operators, e.g.
\be
\frac{\langle {Swap_{A}} \rangle}{\langle {Swap_{A'}} \rangle} =\frac {\sum_{lr} w_r w_r\langle V_l \lvert Swap_{A'} \lvert V_r\rangle\frac{  \langle V_l  \lvert Swap_A \lvert V_r\rangle}{ \langle V_l \lvert Swap_{A'} \lvert V_r\rangle}}
		{\sum_{lr} w_l w_r \langle V_l \lvert Swap_{A'} \lvert V_r\rangle}. \label{ratioT}
\ee
This improves the sampling since, if regions $A$ and $A'$ are similar in size, the measurement $\tfrac{  \langle V_l  \lvert Swap_A \lvert V_r\rangle}{ \langle V_l \lvert Swap_{A'} \lvert V_r\rangle}$ will have fewer possible values than $\tfrac{  \langle V_l  \lvert Swap_A \lvert V_r\rangle}{ \langle V_l \lvert V_r\rangle}$, and those values will have a smaller variance.

Note however that one is only measuring a {\it ratio} of expectation values.
That is, to obtain $ \langle Swap_A \rangle $, one must know the value of $ \langle Swap_{A'} \rangle$.  If  $ \langle Swap_{A'} \rangle$
was also obtained by a ratio trick simulation, the expectation value for the smaller component of $A'$ must be determined, and so on.
Thus, the procedure that we use in this paper is to measure a range of sizes for region $A$, beginning with a measurement of the bare $Swap$ for a small region size, and increase the size of regions $A$ and $A'$ over several simulations in sequence.
That is, we measure {\it Swap} for a sequence of different region sizes, $A_1, A_2, \cdots,  A_n$, where the number of lattice sites in $A_{i+1}$ is greater than the number of sites in region $A_i$.  Then, the Renyi entropy of an arbitrary region $A_n$ is calculated through
\begin{eqnarray}
S_2(A_n) &=& -\ln \left({ \frac{\langle {Swap_{A_n}} \rangle}{\langle {Swap_{A_{n-1}}} \rangle}  }\right) - 
\ln \left({ \frac{\langle {Swap_{A_{n-1}}} \rangle}{\langle {Swap_{A_{n-2}}} \rangle}  }\right) \nonumber \\
&-& \cdots - \ln \left({ \frac{\langle {Swap_{A_2}} \rangle}{\langle {Swap_{A_{1}}} \rangle}  }\right) 
- \ln \left({ \langle {Swap_{A_1}} \rangle }\right) \label{Ssum}
\end{eqnarray}
where each ratio is calculated via Eq.~\ref{ratioT}, and the last expectation value for $A_1$ via Eq.~\ref{weight}.
Note that each term in the sum requires a different VB QMC simulation, since, although
we can measure the entropy for any region $A$ within one simulation, we can only use {\it one} size of  $A'$ per simulation, since it affects the sampling of the valence bond states as described below.  The scaling cost of the Ratio trick is therefore $n$; however, the gain in 
sampling efficiency is demonstrated to more than compensate for this additional simulation cost.

\subsection{The Loop/Ratio Algorithm}

\begin{figure} {
\includegraphics[width=3.0 in]{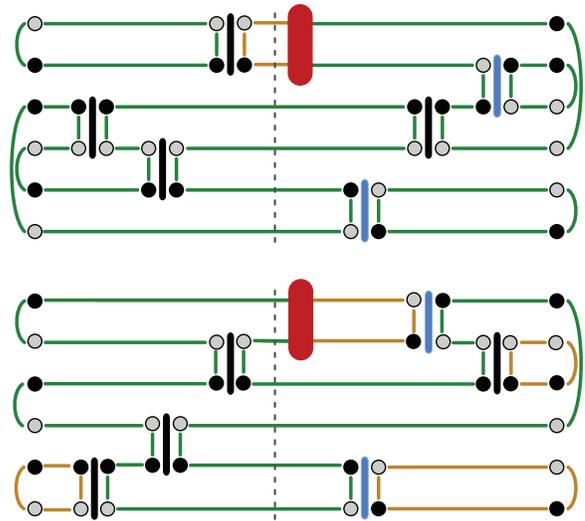} \caption{ 
\label{lratio}
One possible simulation cell configuration for the loop ratio algorithm on a 6-site system where $\alpha=2$ and region $A'$ contains the first 2 sites of the system.
Spins between the usually non-interacting copies are connected through loops via the $Swap$ operator (red). 
The green and orange links are used to show the connections between the sheets.  
The loop on the left side of the swap operator on the top sheet is connected to the right side on the bottom sheet and so forth.
}
} \end{figure}

In order to calculate Renyi entanglement entropy with maximal efficiency in VB QMC simulations of the Heisenberg model, an
algorithm should be employed that combines the loop update with the ratio trick.
When modifying the loop algorithm to use the ratio trick, the same principles as in the double-projector algorithm (above) apply, 
however the sampling weight (from Eq.~\eqref{weight}) is not explicit since one instead samples over spin states whose overlap is always unity.

In order to made the necessary modification to the loop algorithm, the system should first be replicated so that two non-interacting
copies are present, as usual for measurements of the {\it Swap} operator.\cite{swap}
Then, links in the simulation cell are reconnected as if there were a $Swap$ operator permanently applied to the projected state $\lvert V_r \rangle$, shown in Fig.~\ref{lratio}.
This causes spins from different non-interacting copies of the system to be connected via loops, which means they can be flipped together, and thus the spin states are sampled according to the swapped system $\langle V_l  \lvert Swap_{A'} \lvert V_r\rangle$.
The measurement of  $\langle V_l  \lvert Swap_{A} \lvert V_r\rangle/\langle V_l  \lvert Swap_{A'} \lvert V_r\rangle$ is then accomplished by measuring an operator which swaps the states of the sites in region $A$ that were {\it not} already swapped in region $A'$, assuming $A' \subset A$.

This method has the same limitation of the double-projector ratio trick, that only one value of $A'$ can be used per simulation, so the region to be measured must be built up from a small region $A$ according to Eq.~\eqref{Ssum}.  In our results below, we use two geometries for 
building the region $A$: ``strips" and ``squares" (see Fig.~\ref{arealaw} or Ref.~\onlinecite{critMI}).  Strips refer to geometries where region $A$ has one dimension equal to the linear size of the toroidal system itself, and is therefore without corners.  The Renyi entropy $S_2(A)$ is built up through the ratio trick by systematically adding sub-regions of size $L \times 1$.  Squares refer to geometries where the linear size of $A$ is increased symmetrically, starting from size $1 \times 1$.  Square regions $A$ necessarily have four corners.

\subsection{Results}

\begin{figure} {
\includegraphics[width=3.3 in]{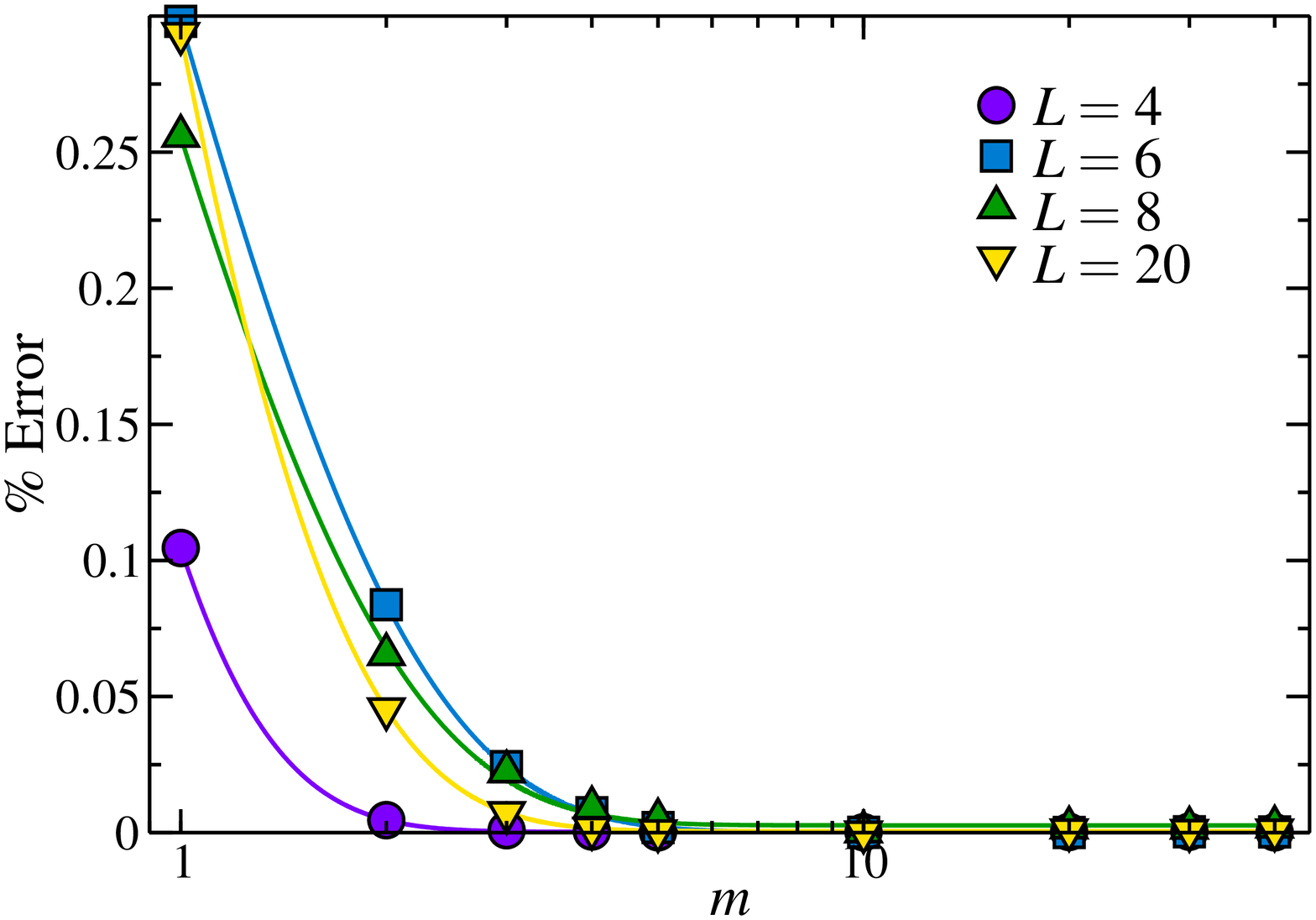} \caption{ 
\label{mconv}
Percent error in $S_2$ versus the number of operators per site ($m$) for different $L\times{L}$ lattices.  
For $L=4,6,8$ the lattices have open boundaries and region $A$ is half the system using the strip geometry.  The exact values were found using density matrix renormalization group (DMRG) simulations.
The $L=20$ lattice has periodic boundaries and $A$ is a $2\times2$ square.  The ``exact'' value is taken from the $m=10$ simulation. 
Each data set was fit to an exponential function.
}
} \end{figure}

\begin{figure} {
\includegraphics[width=3.3 in]{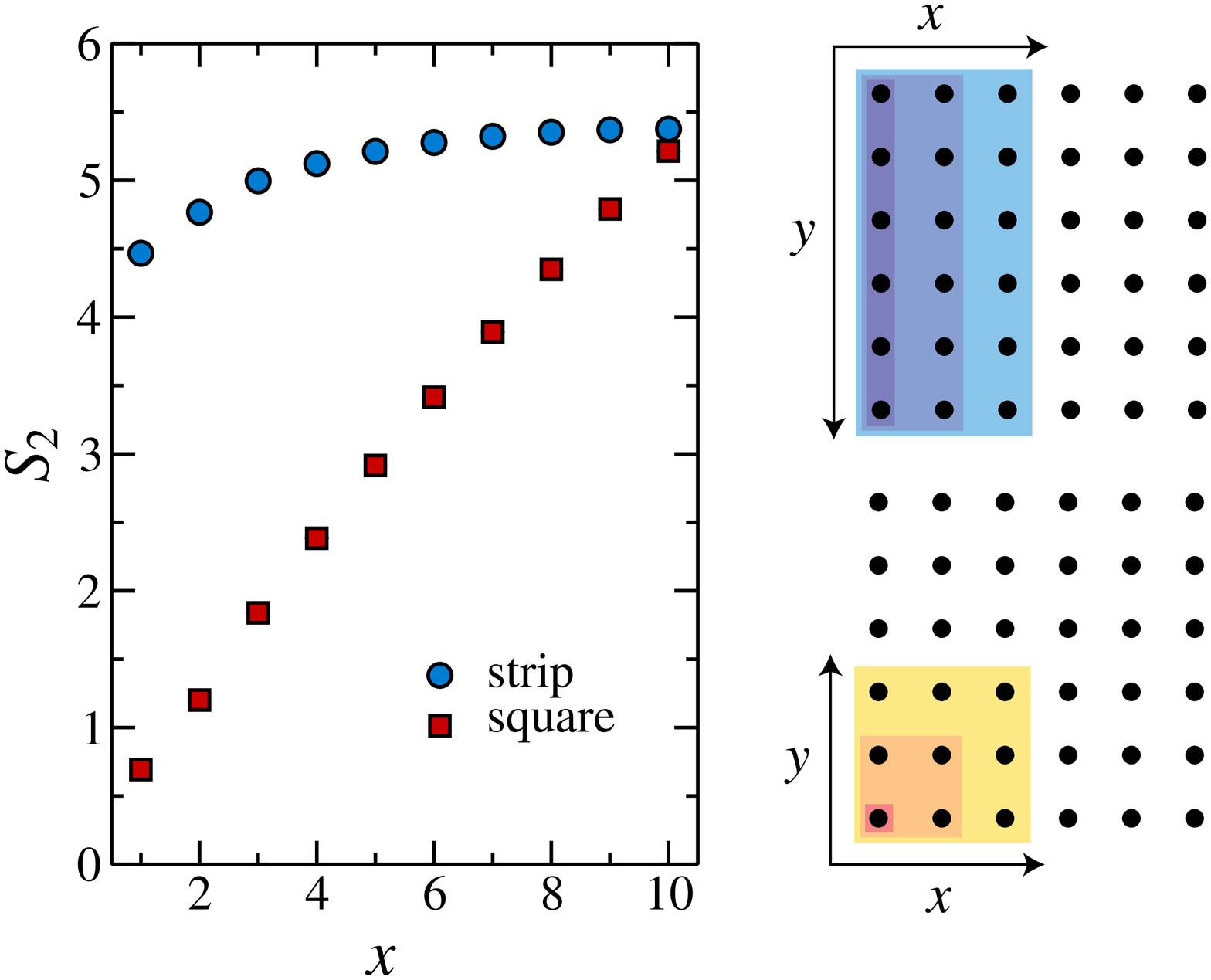} \caption{ 
\label{arealaw}
(Left) The Renyi entropy for a $20\times20$ PBC system versus the width ($x$) of region $A$ for both the square and strip geometries.  
The boundary length does not change with the region width for strip geometry (since its height $y$ traverses the periodic lattice), thus the entropy approaches a constant value.
For square geometry the boundary length is $4x$.
In both geometries the entropy scales with the length of the boundary, that scaling becoming better as the width of region $A$ approaches half the system size.
At right, the ``strip" (top) and ``square" (bottom) geometries on a $6\times6$ lattice.
}
} \end{figure}

We begin by testing some basic properties of the Renyi entropy as calculated through the VB basis QMC with the loop/ratio trick outlined above.  First, we examine the convergence of the Renyi entropy as a function of operator list length per site, $m=M/N$, as illustrated in Fig.~\ref{mconv}.  
Using several system sizes, both periodic and open boundaries, and strip and square geometries (described above and in Fig.~\ref{arealaw}), we see very good convergence by $m=10$.
This value of $m$ was used for all the following VB QMC simulations, as the measurements are able to converge with that number of operators, but additional operators would detract from the algorithm's efficiency.

Figure \ref{arealaw} shows examples of $S_2(A)$ for regions $A$ of both strip and square geometries of increasing width 
for a $20\times20$ toroidal system.
The length of the boundary for all strip regions in this plot is $\ell = 40$, whereas for the square regions $\ell = 4x$.
The area law scaling of $S_2$ is apparent in that the strip and square geometries both approach a straight line with zero and non-zero slope respectively.

\begin{figure} {
\includegraphics[width=3.3 in]{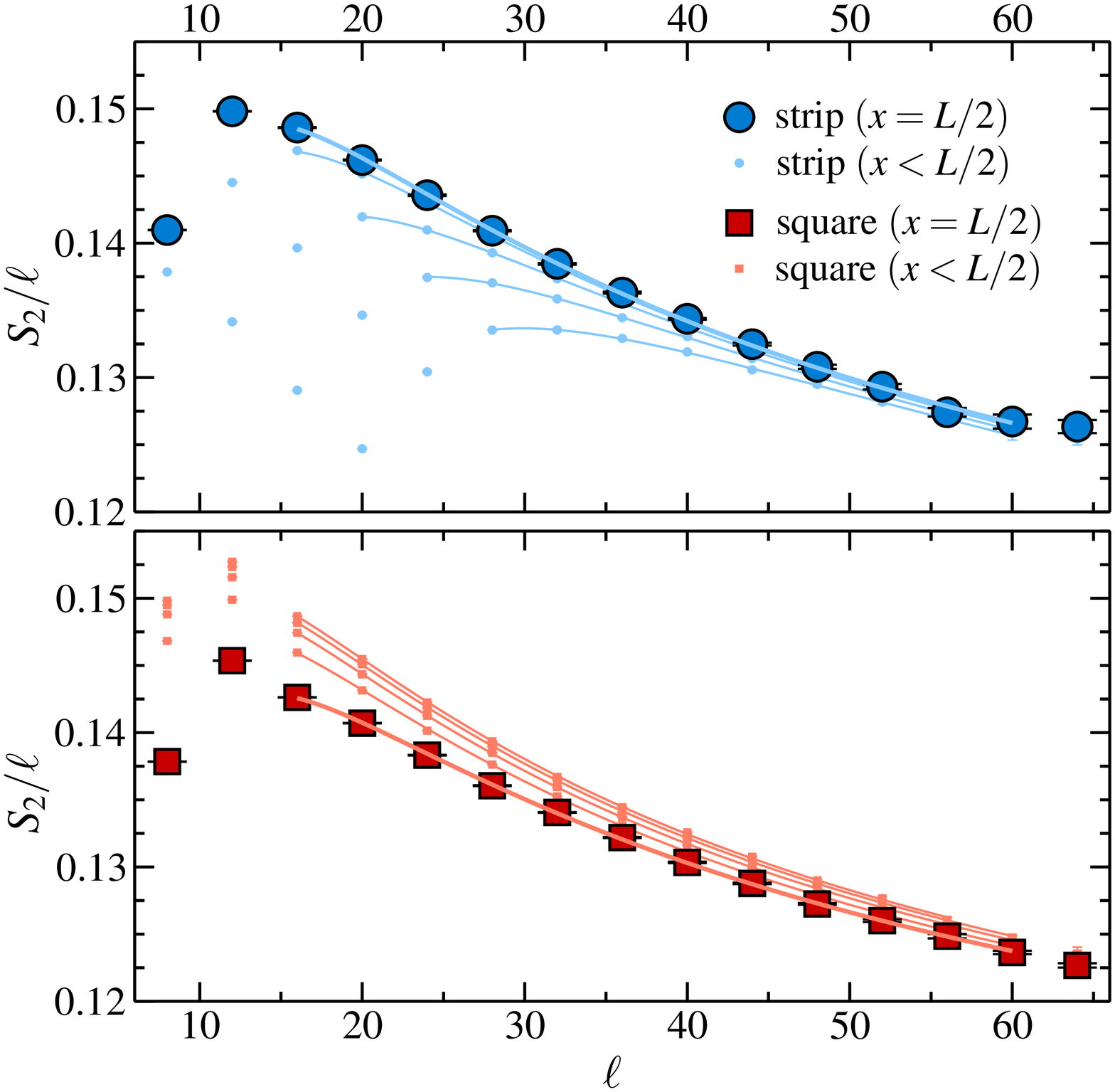} \caption{ 
\label{qmcscaling}
$S_2/\ell$ vs $\ell$, where $\ell$ is the boundary length of region $A$, for regions with square and strip geometry 
embedded in $L\times{L}$ systems with periodic boundary conditions.
The data sets outlined in black correspond to region $A$ with width $x=L/2$ for both geometries.
The other data sets use smaller region $A$ of the same geometries.
Fits to Eq.~\eqref{scalingeq} are included for all data, and the coefficients found are listed in Table~\ref{coeffs}.}
} \end{figure}

To determine the scaling of entanglement entropy in two dimensions, we examine $L\times{L}$ systems with periodic boundary conditions, the results of which are shown in Fig.~\ref{qmcscaling}.
Region $A$ was systematically built up according to the square and strip geometries as defined in Fig.~\ref{arealaw}.
Since each region is built to satisfy Eq.~\eqref{Ssum}, one can perform fits for several sizes of subregion for each lattice size.
In this case (as opposed to the plot in Fig.~\ref{arealaw}), for each set of data in Fig.~\ref{qmcscaling} we use a region $A$ with width proportional to the system size.  This is done in attempt to overcome finite size effects, and the interaction of boundaries (that can be seen for the strip geometry in Fig.~\ref{arealaw}).
Fig.~\ref{qmcscaling} includes data for $S_2(A)$ using regions with width $x=L/2$ as well as smaller regions $x<L/2$.
A smaller region $A$ has the advantage that $S_2(A)$ converges faster, but the drawback is it brings one into the regime of finite-size effects,
apparent in Fig.~\ref{arealaw} with small $x$, where $S_2$ is lower than the area law value found at $x=L/2$.
This deviation seems to depend on the fraction of the system contained within region $A$ which is why, for the square case (where both the size of region $A$ and the boundary length are scaled with system size), the different region widths do not change the entropy scaling very much.
However, for the strip geometry with a region $A$ width of $x=L/2-n$, the fraction of the system contained in region $A$ changes with L, and approaches $1/2$ as L increases, i.e. when $L/2 \gg n$.
This effect is evident in the top panel of Fig.~\ref{qmcscaling}, where for $n \ne 0$ the entropy is diminished for smaller system sizes, but approaches the $n = 0$ values as system size increases.

As is clear, the data gives excellent fits to the function
\be
\frac{f(\ell)}{\ell} = a + \frac{c}{\ell}\ln(\ell) + \frac{d}{\ell}, \label{scalingeq}
\ee
where $\ell$ is the length of the boundary between regions $A$ and $B$, provided that the very smallest lattice sizes are excluded.
The values obtained for coefficients $a, c,$ and $d$ are listed in Table \ref{coeffs}.

\begin{table}[ht]
  \begin{tabular}{| l | c || l | l | l | }
  \hline
    geometry &$L/2-x$ & $\:\:\:\:\:\;a$ & $\;\;\;\;c$ & $\;\,-d$ \\ \hline \hline
        strip & 0 & 0.0965(7) & 0.74(2) & 1.22(5) \\ \hline
                 & 1 & 0.0956(6) & 0.79(2) & 1.36(4) \\ \hline
                 & 2 & 0.0923(8) & 0.96(3) & 1.88(6) \\ \hline
                 & 3 & 0.088(1) & 1.20(5) & 2.6(1) \\ \hline
                 & 4 & 0.082(1) & 1.54(6) & 3.7(2) \\ \hline
        square &   0 & 0.0976(3) & 0.64(1) & 1.06(2) \\ \hline
                      & 1 & 0.0976(3) & 0.621(9) & 0.95(2) \\ \hline
                      & 2 & 0.0977(2) & 0.617(8) & 0.91(1) \\ \hline
                      & 3 & 0.0976(2) & 0.626(8) & 0.93(2) \\ \hline
                      & 4 & 0.0975(5) & 0.63(1) & 0.94(3) \\ \hline
 \end{tabular}
   \caption{\label{coeffs}
   The coefficients $a$, $c$, $d$ found by fitting the data in Fig.~\ref{qmcscaling} to Eq.~\eqref{scalingeq}.
   The fits were done for both strip and square geometries, beginning at $x=L/2$ for both and decreasing the region sizes until $x=L/2-4$, where $x$ is defined in Fig.~\ref{arealaw} as the width of region $A$.
   }
 \end{table}

\section{Series Expansions}  \label{HTEsec}

\subsection{Expansion Methods}
We have developed two different types of series expansions to calculate
the entanglement properties of the Heisenberg model. The first is High Temperature
Expansions (HTE). This method was introduced in Ref.~[\onlinecite{XXZ}] for the XXZ model,
of which the Heisenberg model is a special case. In this method, the
mutual information between two regions $A$ and $B$ associated with their
boundaries or corners can be expanded in powers of inverse temperature $\beta$. The calculation can be done for Renyi mutual information of index $\alpha$ by
introducing $\alpha$ replicas of the system. The coefficient of $\beta^\alpha$ is a
polynomial in $\alpha$ of order $\alpha-1$ so that the limit $\alpha \to 1$ can be readily
taken to calculate the von Neumann mutual information as well.

Since there is no finite temperature phase transition in the Heisenberg
model on the square lattice, the expansions in $\beta$ can, in principle,
be extrapolated down to $T=0$. It is well known that the correlation
length of the system grows exponentially at low temperatures as
$\exp{(C/T)}$. Hence, we expect corrections to leading behavior to be
exponentially small at low temperature, $\exp{(-C/T)}$. For this reason,
a change of variables $w=\tanh{\beta}$ is applied before Pade approximants
are used. This extrapolation method has been used in the past for other
properties of the Heisenberg model\cite{singh-gelfand} and also shows good convergence
between different Pade approximants when applied to mutual information.

In addition, we have developed a series expansion directly for the
entanglement entropy at $T=0$ using Ising anisotropy parameter $\lambda=
J_{xy}/J_z$ of the XXZ model. When $\lambda=0$, we pick one of
the two Ising states to expand around. That unique state factorizes
for any two regions $A$ and $B$. So, any entanglement entropy vanishes
in that limit. For any $\alpha \ge 2$, the series expansion for the Renyi
entropy can be calculated as a power series in $\lambda$ using a
linked cluster expansion.\cite{gelfand-singh-huse,oitmaa} 
Unlike HTE, the coefficients of the
Renyi entropies for different $\alpha$ are not related by a simple polynomial
relation and for every $\alpha$ a different calculation is needed. Here,
we will restrict ourselves to the calculation of Renyi entropy with
$\alpha=2$.

The Ising series expansions work in the thermodynamic limit, starting
with a system that has a short correlation length and long-range order
along a particular direction. Since a specific ordered state is picked out,
their major limitation is that they cannot be used to study
any bulk entanglement properties associated with broken symmetry. 
In any finite order of perturbation theory,
contributions can come only from the boundary between regions $A$ and $B$.
The series are non-singular as long as the system has a gap.
The singular dependence on the size $L$ of the system is replaced 
in the series expansion studies by
a dependence on the correlation length $\xi$, which diverges
as the gapless Heisenberg point is approached.
By scaling, the dependence on $L$ should translate into
a similar dependence on $\xi$.
An advantage of series expansions is
that entanglement associated with different surface manifolds
such as surfaces, lines and corners can be analytically separated
and separate expansions can be obtained for the entropy
associated with them.

\subsection{Results from Series Expansions}

\begin{figure} {
\includegraphics[width=3.3 in]{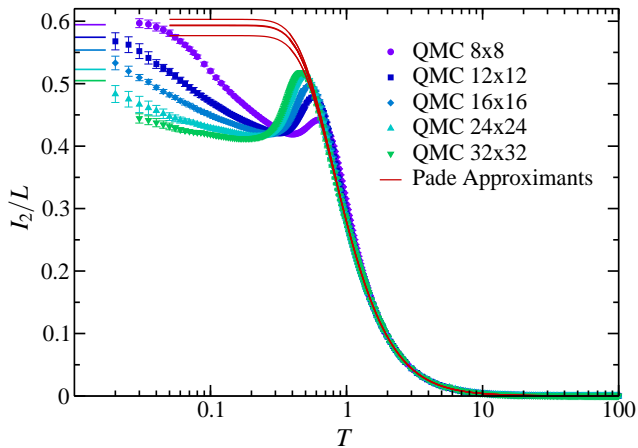} \caption{ 
\label{pade}
The Finite temperature Renyi mutual information for $L\times{L}$ periodic systems with strip geometry (note $\ell=2L$) for region $A$, with SSE QMC results for $L=8,12,16,24,32$, and [4/6], [5/5], [6/5] and [6/4]
Pade approximants for the HTE in the variable $w=\tanh{\beta}$.
The horizontal lines on the left hand side correspond to the zero temperature results (from VB QMC) for each of the finite temperature SSE data sets.
}
} \end{figure}

\begin{figure} {
\includegraphics[width=3.1in]{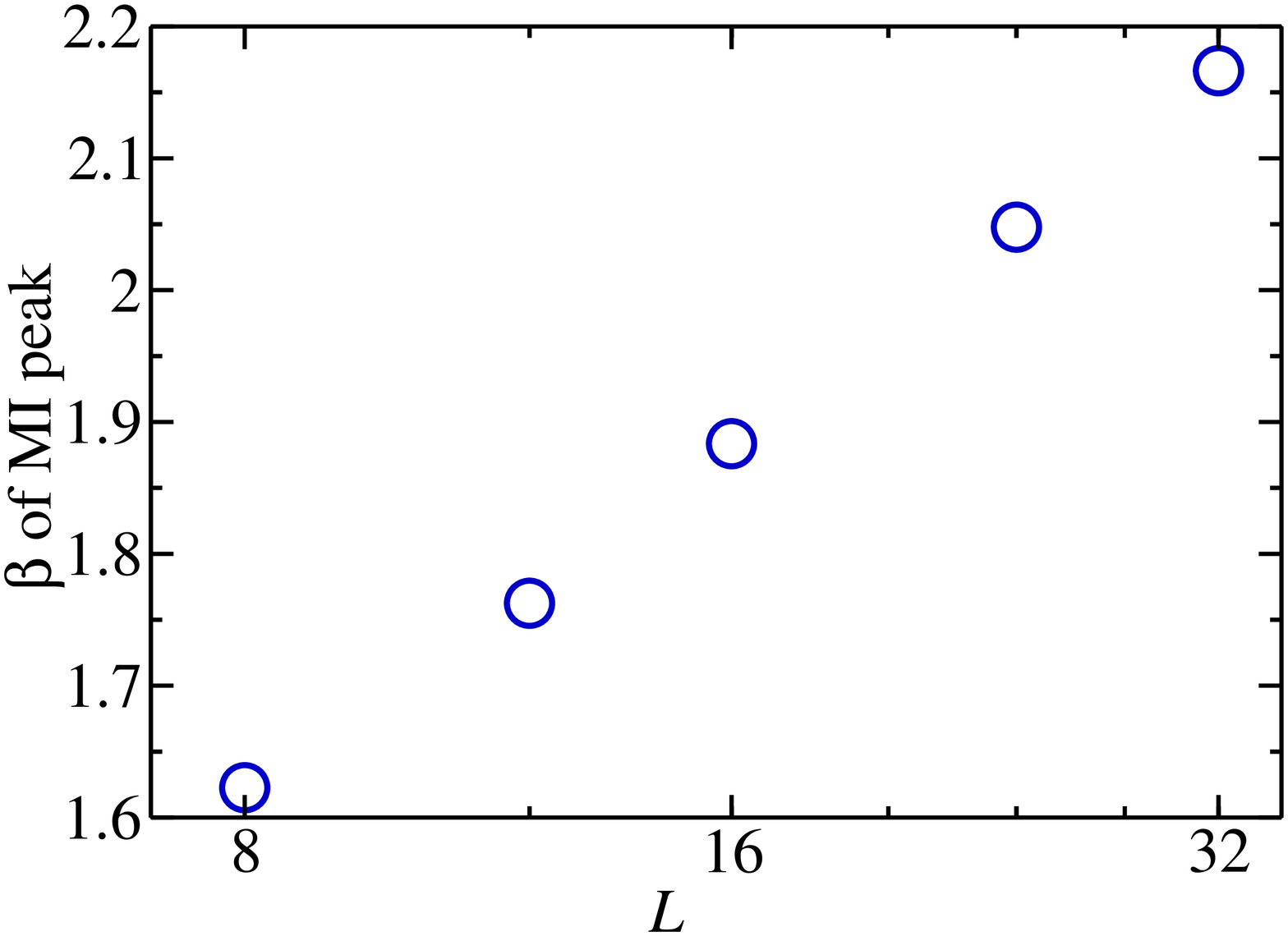} 
\caption{ 
\label{crossover}
The inverse {crossover} temperature of Renyi mutual information for the system sizes shown in Fig.~\ref{pade}, where the crossover temperature is taken as the temperature at the highest point of the $T>0$ mutual information peak.
The x axis is logarithmic, showing that the crossover temperature scales as $\sim{\log(L)}$.
}
} \end{figure}

Pade extrapolation of High Temperature Expansions (HTE) in the variable
$w=\tanh{\beta}$ are compared with the data from stochastic series expansion (SSE) quantum Monte Carlo (QMC)
simulations in Fig.~\ref{pade}. The agreement is excellent
down to temperatures below $J$. At low temperatures the QMC data
shows dramatic finite-size effects, with a maximum and a minimum,
followed by a slow rise at lower temperatures. In contrast, the
Pade approximants show a steady monotonic rise and saturation at
low temperatures. These results suggest that the limit of $T\to 0$ 
and $L\to\infty$ do not commute for these quantities. We note
that such non-commuting limits are well known for other
properties of the Heisenberg model.\cite{DSFisher,Hasenfratz} However, these non-commuting limits
have not been anticipated for the mutual information.  In Fig.~\ref{crossover},
we plot the crossover temperature, as defined by the temperature of the local maximum in mutual information at $T>0$,
as a function of $L$.  We see that the crossover temperature scales inversely with $\log(L)$, or, equivalently,
$L$ scales exponentially in the inverse temperature.  Such a length scale agrees with the correlation length in the Heisenberg
model, which also scales exponentially at nonzero temperature, and so at first sight it is natural to ascribe this peak in the mutual information
to the size of the correlation domains.  One might imagine the following argument: when the size of the system is larger than
the correlation domain size, there are many correlation domains along the boundary.  Since the direction of the spin in a correlation
domain can be viewed as roughly uncorrelated with that in other correlation domains, this contributes an additional term to the entropy of both regions.  However, the correlation domains that cross the boundary produce some correlations between the two regions, leading to a positive contribution to the mutual information.  
While this argument captures the correct qualitative scaling, suggesting that the mutual information should drop at lower temperatures, it fails on quantitative grounds.  The contribution to the entropy of a correlation domain from the random ordering direction of that domain should scale something like the logarithm of the correlation volume (see, for example, the mean-field theory calculation of the next section, where at low temperatures the entire system comprises one correlation domain), and hence should be proportional to $\beta$.  However, the density of these correlation domains (the number of domains per unit boundary length, which is inversely proportional to the correlation length) is exponentially small in $\beta$, and so we should expect that this contribution to the entropy per unit length should become negligible as $\ell$ gets larger.  In contrast, the numerical data shows the difference between the $T>0$ maximum and the $T=0$ limit increasing with increasing $\ell$, so that this difference remains as an unexplained phenomenon.

Let the expansions around the Ising limit for the second Renyi entropy
per unit length of the boundary be given by
$$
S_{2l}/\ell = \sum_{n=2} a_n \lambda^n,
$$
and, for a single corner,
$$
S_{2c} = \sum_{n=4} b_n \lambda^n.
$$
The non-zero coefficients up to $n=14$ are given in Table~2.
Note that all odd order terms are zero, so the expansion is in
the variable $\lambda^2$.
The series for the line term $S_{2l}$ are evaluated by first
performing a change of variables $\delta=\lambda^2-2\lambda^4$
to remove a square root singularity at $\lambda=1$ and then
calculating Pade approximants. From these we estimate
$$
S_{2l}/\ell=0.094\pm 0.001,
$$
for the Heisenberg model. Here, the error bars represent
spread in values obtained
from different Pade approximants.\cite{oitmaa} For the corner term, we expect a
logarithmic singularity of the form
$$
S_{2c}=x\ln\xi\sim -{x\over 2}\ln{(1-\lambda^2)},
$$
where, we have used the fact that the correlation length
diverges as $(1-\lambda^2)^{-1/2}$. Given the anticipated
logarithmic singularity, we first differentiate the 
series with respect to the variable $\lambda^2$, and then
use Pade approximants biased at $\lambda^2=1$ to estimate
the residue. From these, we estimate
$$
x=-0.020\pm 0.002.
$$
Numerical values for $S_{2l}/\ell$ and $x$ will be compared to VB QMC results and
other calculations later in the discussion section.

\begin{table}
  \begin{tabular}{| l | c | c | }
  \hline
    $n$ &  $a_n$     & $b_n$ \\ \hline \hline
    2 & 0.05555556 &  0    \\ \hline
    4 & 0.00314815 &  0.00913580 \\ \hline
    6 & 0.00558342 & -0.00542753 \\ \hline
    8 & 0.00353554 & -0.00126847 \\ \hline
   10 & 0.00220329 & -0.00172570 \\ \hline
   12 & 0.00186784 & -0.00170418 \\ \hline
   14 & 0.00144690 & -0.00144091 \\ \hline
 \end{tabular}
   \caption{\label{series-coeffs}
Ising series expansion coefficients for the line and corner terms
   }
\end{table}

\section{Mean-Field Theory and Tower of States Modes}
\label{sec:mft}
\subsection{Tower of States and Thermodynamics}
One starting point for a theoretical treatment of the Heisenberg model is a mean-field theory.  This mean-field theory provides a 
simple framework for understanding some logarithmic bulk corrections associated with spontaneous symmetry breaking, as well as providing some understanding
of the low energy ``tower of states" modes (first identified by Anderson\cite{tower}) and which are present even in the two dimensional model.

In mean-field theory we consider the model Hamiltonian
\be
H=\frac{J}{N} \sum_{i \in 1, j \in 2} \vec{S_i} \cdot \vec{S_j},
\ee
where the sum ranges over spin $i$ in one sublattice and $j$ in the other sublattice (we use $1$ and $2$ to denote sublattices), and
there are total of $N$ spins,  with $N/2$ spins in each sublattice.  The ground state of this Hamiltonian corresponds to taking all spins in sublattice
$1$ in a symmetric state, with total spin $N/4$, and similarly taking all spins in sublattice $2$ in a symmetric state, and then pair the
two spins to form a singlet.

We begin by working out thermodynamic properties of the model, and discuss their impact on QMC simulations using SSE and VB projector methods.
We then turn to the question of the entanglement entropy.

First, we work out the energy of the aforementioned singlet state as follows: let $S_1$
denote the total spin operator in sublattice $1$, and $S_2$ denote the
same for sublattice 2.  Then, the Hamiltonian can be written as $(J/N)
S_1 \cdot S_2$.   This equals $(J/2N) [ (S_1+S_2)^2-S_1^2-S_2^2]$.  
The
singlet state has $S_1+S_2=0$.  Looking at states with spin $N/4$ in each
sublattice, we have $S_1^2=S_2^2=(N/4)(N/4+1)$, so, the energy is
$-(J/N) (N/4) (N/4+1)$, which is of order $N$.  

One can now look at excited
states.  The excited states where both sublattices have total spin
$N/4$, but the whole system is not in a singlet, give the ``tower states":
we find that in this case $S_1+S_2$ is not equal to zero.  Indeed, the
energy difference of this state, compared to the ground state, is
simply given by 
\be
E-E_{ground}=(J/2N) (S_1+S_2)^2.
\ee
That is, the energy is
equal to $(J/2N)$ times the total spin squared.  In real two-dimensional systems,
a similar low energy structure of states is observed, with the energy of
these states proportional to total spin squared divided by $N$, although the
coupling constant $J$ describing the energy of these
modes may be renormalized compared to the coupling constant appearing in the
lattice Hamiltonian.
These low energy states can be observed in exact diagonalization, and in fact
they are one of the best checks for the presence of symmetry breaking.\cite{tower2}
Note that these states are much lower energy than the spin-wave states: in a
2D system with linear size $L$, they have energy of order $1/L^2$, while
the lowest energy spin-wave has an energy of order $1/L$.

There are also excited states where a given sublattice does not have
total spin $N/4$.  That is, not all spins in the same sublattice are in
a symmetric state.  One can check in this case that the energy of
these states is increased above the ground state energy by an amount
that is of order $1$ (or more, if the spin is reduced a lot compared to
$N/4$).  These states can be viewed as a mean-field theory analogue of the
spin-wave states; that is, the mean-field theory raises the energy of the
spin-waves from order $1/L$ to order $1$.

So, at a temperature of order $1$, it becomes reasonable to
ignore those states in mean-field theory (more precisely we need a temp of order
$1/\ln(N)$).  In a two-dimensional system, the temperature needs to become of order $1/L$
to ignore the spin-wave modes.

We now consider the effect of the tower of states on the bulk entropy, at temperature
sufficiently low that the spin-waves can be ignored.  We emphasize that this estimate
is {\it not} a calculation of the entanglement entropy, but rather a calculation of
a bulk thermodynamic entropy.
Considering the tower states, the total entropy is not
hard to work out: the allowed states are $1$ state (the ground state) which is a singlet, $3$
states with spin $1$ and energy increased by $J S(S+1)/2N=2J/2N$, $5$ states
with spin $2$ and energy $6J/2N$, etc... (in this case, the
Clebsch-Gordon coefficients work out simply, so that the number of states
with spin $S$ is exactly $2S+1$).  As a rough approximation, at a
temperature $T$, we expect to excite states with spin such that $JS^2/N$ is
of order $T$.  So, $S$ is of order $\sqrt{TN/J}$ or less.  The number of such
states is of order $S^2$, and hence of order $TN/J$.  Thus, at a temperature $T$ much larger than $1/N$,
but sufficiently small that the spin-waves can be ignored, the
entropy is equal to
\be
\log(TN/J)+{\rm const.}
\ee

\subsection{Effect of Tower of States on QMC Simulations}
In QMC simulations, if we want to access the ground state, it is important to access a temperature sufficiently small
that these tower of states modes can be ignored.  Since the energy of the tower of states modes is so small (of order
$1/N$) this can require a prohibitively small temperature.  However, we will see that this is {\it not} as big a problem
as it might seem.  

First, consider the VB projector method.  The projector method starts with a trial wavefunction
and then applies a large power of the Hamiltonian to this wavefunction.  The Hamiltonian is appropriately scaled such that
its ground state will have the largest eigenvalue, so this high power of the Hamiltonian, acting on the trial wavefunction,
produces a state close to the ground state.  This is essentially the power method of finding largest or smallest eigenvalues of
a matrix.  Given that the Hamiltonian has tower of states modes with energy very close to the ground state, it would seem that
we would have to apply a very high power of the Hamiltonian, a power which is of order $N$, in order to produce a final state
close to the ground state.  However, the trial wavefunction in the VB projector method has total spin $0$, and the Hamiltonian
conserves spin.  Thus, the wavefunction produced by acting on the trial wavefunction with a high power of the Hamiltonian also
has total spin $0$.  However, all the tower of states modes have non-zero spin, and hence the wavefunction we produce has no overlap
with the tower of states modes.  This is the reason why it suffices to simply go to high enough order to project out the spin-wave modes,
as once the spin-waves excitations are projected out, the tower of states are also projected out by symmetry (note that we do have
excited states with  total spin $0$ which excite both a spin-wave and a tower mode, but such states have energy of order $1/L$ or higher,
not $1/N$).

Consider finally the SSE method.  As discussed above, the tower of states modes do have a noticeable effect on the calculation of the
bulk entropy.  However, as we will see below, at least in mean-field theory they have only a small effect on the calculation of the entropy of the reduced density matrix, suggesting that the SSE calculations of the reduced density matrix entropy converge well even without accessing temperature of order $1/N$.

\subsection{Entanglement Entropy}
We now consider the entanglement entropy in the mean-field model.  The  entanglement between the $1$ and $2$ sublattices is the simplest
to calculate.  Each sublattice has total spin $N/4$, and hence has $N/2+1$ states.  The ground state, which we call $\psi^0$, is
maximally entangled, and hence has entanglement entropy $\ln(N/2+1)$, with all different Renyi entropies equal.

The more interesting entanglement entropy to calculate is to imagine dividing the system into two halves, $A$ and $B$, with
each half having $N/4$ spins in sublattice $1$ and $N/4$ spins in sublattice $2$.  This is, in our opinion, the simplest mean-field
model which is relevant to the numerical calculations on the two dimensional Heisenberg model, as in that case each region $A$ and $B$
contains spins from both sublattices.  We will see that this calculation gives a logarithmic dependence on $N$ also, although the result is more complicated.

The ground state $\psi^0$ can be obtained by taking {\it any} state which has spin $N/4$ in each sublattice and which has
non-zero overlap with the ground state and projecting it into the
total spin zero sector.  We choose to use a N\'eel state as our state before projection, where we define this state $\psi^{\rm{N\acute{e}el}}$ as a state in
which all spins in sublattice $1$ are pointing up and all spins in sublattice $2$ are pointing down.
We do this projection by averaging over different rotations of the N\'eel state,
so
\be
\psi^0=Z^{-1/2} \int {\rm d}\theta
{\rm d}\phi R(\theta,\phi) \psi^{\rm{N\acute{e}el}},
\ee
where $R(\theta,\phi)$ is the rotation by
angles $\theta$ and $\phi$, and the measure is chosen to be uniform over all rotations (that is,
we choose the Haar measure), and where $Z$ is a normalization factor so that $|\psi^0|^2=1$.

We compute the second Renyi entropy, $S_2$.  The calculation of other Renyi entropies are similar; in this case, in contrast to
the previous entropy calculation above, different Renyi entropies differ,
so that $S_2$ is not equal to the von Neumann entropy.
To calculate the Renyi entropy, we must calculate
\be
\langle \psi^0 \otimes \psi^0 | Swap_A | \psi^0 \otimes \psi^0 \rangle,
\ee
where $Swap_A$ is the swap operator used in Ref.~[\onlinecite{swap}].

This expectation value is equal to
\begin{eqnarray}
\label{swapval}
&&Z^{-2} \int {\rm d}\theta_1 {\rm d}\phi_1 {\rm d}\theta_2 {\rm d}\phi_2 {\rm d}\theta_3 {\rm d}\phi_3 {\rm d}\theta_4 {\rm d}\phi_4
\\ \nonumber
&\times&\langle \psi^{\rm{N\acute{e}el}} \otimes \psi^{\rm{N\acute{e}el}}|R(\theta_1,\phi_1)^\dagger \otimes R(\theta_2,\phi_2)^\dagger
| Swap_A | 
\\ \nonumber
&& R(\theta_3,\phi_3) \otimes R(\theta_4,\phi_4) | \psi^{\rm{N\acute{e}el}} \otimes \psi^{\rm{N\acute{e}el}}\rangle.
\end{eqnarray}

We first estimate $Z$ as follows.  We have
\be
Z=
\int {\rm d}\theta_1 {\rm d}\phi_1 
{\rm d}\theta_2 {\rm d}\phi_2
\langle \psi^{\rm{N\acute{e}el}} | R(\theta_1,\phi_1)^\dagger R(\theta_2,\phi_2) | \psi^{\rm{N\acute{e}el}} \rangle.
\ee
We can combine the rotations
$R(\theta_1,\phi_1)^\dagger R(\theta_2,\phi_2)$ into one rotation by a pair of combined angles, $R(\theta,\phi)$.  Equivalently,
we note that the integral
$\int {\rm d}\theta_2 {\rm d}\phi_2
\langle \psi^{\rm{N\acute{e}el}} | R(\theta_1,\phi_1)^\dagger R(\theta_2,\phi_2) |\psi^{\rm{N\acute{e}el}}  \rangle$ is independent of  $\theta_1,\phi_1$, so we
can fix $\theta_1=\phi_1=0$.
Thus, up to constant factors, we have
\be 
Z =\int {\rm d}\phi {\rm d}\theta \langle\psi^{\rm{N\acute{e}el}}  | R(\theta,\phi) | \psi^{\rm{N\acute{e}el}} \rangle. 
\ee

The expectation value in the above integral is just the $N$-th power of
of $\langle \uparrow \rvert R(\theta,\phi) \lvert \uparrow \rangle$.  This is approximated
by (for small $\theta,\phi$) $\exp(-N (\theta^2+\phi^2))$.  Because of the factor of $N$ in the
exponent, the restriction to small $\theta,\phi$ is justified as the expectation value is
negligible for large $\theta,\phi$.  Then, the integral over $\theta,\phi$ is Gaussian and
the result is that
\be
Z\propto N^{-1}.
\ee

Now we estimate the integral in Eq.~(\ref{swapval}).
{In this case}, we have an integral over $4$ pairs of angles.  
For the spins not in region $A$ 
the expression is small unless $\theta_1$ is close to $\theta_3$ and $\phi_1$ is close to $\phi_3$ 
and also $\theta_2$ is close to $\theta_4$ and $\phi_2$ is close to $\phi_4$.  
Similarly, for the spins in region $A$ we need
$\theta_1$ close to $\theta_4$ (and $\phi_1$ close to $\phi_4$) and $\theta_2$ close
to $\theta_3$ (and $\phi_2$ close to $\phi_3$).  Thus, the overlap of spins in region $A$ forces certain pairs of angles to be close
and the overlap of spins not in region $A$ forces other pairs of angles to be close.
So, in fact all the angles
need to be very similar.  So, we get (approximately) a Gaussian
integral over three pairs of relative angles, and one overall rotation that we can factor out.  
The result is proportional to $1/N^3$, then, up to constant factors.

Thus, the expectation is
$1/N^3/(1/N^2)=1/N$, giving again an entropy which scales as $\ln(N)+{\rm const.}$ for large $N$.
Finally, we can consider this entropy at a non-zero temperature.  The effect of a non-zero temperature (high enough to excite the tower modes but low enough to avoid exciting spin-wave states so that the total spin in each sublattice will still equal $N/4$) is to 
give a global density matrix
\begin{eqnarray}
\rho&=&Z^{-1} \int {\rm d}\theta_1
{\rm d} \phi_1 {\rm d}\theta_2 {\rm d}\phi_2 F(\theta_1,\phi_1,\theta_2,\phi_2)
\\ \nonumber
&&\times  R(\theta_1,\phi_1) |\psi^{\rm{N\acute{e}el}}\rangle\langle \psi^{\rm{N\acute{e}el}}|
R(\theta_2,\phi_2)^\dagger,
\end{eqnarray}
for some function $F$, which depends only on the relative angles between the two rotations.  That is, at higher temperatures the rotation angles in the
bra and ket vectors become coupled, while at zero temperature $F$ is a constant.
One can check that this still leaves us with an entropy of region $A$ which is equal $\ln(N)+{\rm const.}$, so that, at least in mean-field theory, the
$\ln(N)$ term in the entropy of region $A$ does not depend upon  whether or not the tower modes are excited.

\section{Spin-Wave Theory} \label{SWTsec}

In understanding the area law for the Heisenberg model, an immediate question arises: what is the entanglement entropy in
spin-wave theory?  After all, spin-waves are gapless, and gapless modes might make one concerned whether or not an area
law holds.  In fact, the entanglement entropy of gapless modes depends strongly upon dimension.  Free bosons with a linear
dispersion relation have a logarithmically divergent entanglement entropy in one dimension, following conformal field theory,\cite{Cardy,1dbosecft} but in two or more dimensions they obey an area law.\cite{2Dboson}

In Ref.~\onlinecite{heisenberg1}, it was shown numerically that the Heisenberg model itself obeyed an area law using density matrix renormalization group and
a spin-wave calculation also led to an area law.  In Ref.~\onlinecite{yalegroup}, a spin-wave calculation was carried out for a finite size system and
was shown to roughly match the qualitative behavior from a quantum Monte Carlo simulation.\cite{swap}

While gapless bosons with linear dispersion do produce an area law in two dimensions, they also produce nontrivial exponents
associated with corners.\cite{logcorner}  The entanglement entropy of a region $A$ has a term equal to the log of the length scale of $A$, multiplied by
the sum over corners of a scaling function of angle of each corner.  For the entropy $S_2$, this term is equal to
\be
\approx -0.0062 \ln(\ell)
\ee
for each $90$-degree corner for a real scalar field with linear dispersion.\cite{logcorner}  Note that the sign of this correction is negative.

For the system we are concerned with, we must multiply this result by two.  There are two ways to understand this counting of modes, to see why the result must be multiplied by two.  On the one hand, we can consider an $O(3)$ nonlinear sigma model.  In $2+1$ dimensions, this model
has a symmetry broken phase, and the Heisenberg model ground state corresponds to this phase.  In the symmetry broken phase, there are
two Goldstone modes, corresponding to two different transverse directions in which the order parameter can move.  In the Hamiltonian spin-wave
language, this factor of two again arises, but for a reason which initially might seem to be different.  Suppose we use a spin-wave representation
in which the operator $b^\dagger_i$ always creates an excitation on site $i$ (we are choosing to follow the notation of Ref.~\onlinecite{yalegroup}, though
of course the Hamiltonian spin-wave calculation is a textbook calculation); that is, if we do a spin wave expansion about a state with spins up on the $1$ sublattice and down on the $2$ sublattice, then this operator $b^\dagger_i$ corresponds to a lowering operator on the $1$ sublattice and a raising operator on the $2$ sublattice.  Then, we find gapless modes at momenta near $(0,0)$ and $(\pi,\pi)$ on a two dimensional square lattice.  Thus, we again see a factor of two, arising from the existence of two different gapless points.

We can clarify the relation between the factor of two in these two different approaches.
These gapless modes in the Hamiltonian model correspond to the following states, respectively.  Acting with the operator $\sum_i b^\dagger_i$ on the ground state, which creates a zero energy excitation with momentum $(0,0)$, produces a state which is a superposition of all possible ways of flipping one spin.  Acting with the operator $\sum_i (-1)^i b^\dagger_i$, which
 creates a zero energy excitation with momentum $(\pi,\pi)$, produces a state which again is a superposition of all possible ways of flipping one spin, but with a plus sign for flipping a spin in the $1$ sublattice and a minus sign for flipping a spin in the $2$ sublattice.  
The first of these states corresponds to acting with the operator $\sum_i \sigma^x_i$ on the ground state and the second to acting with the operator $\sum_i \sigma^y_i$ on the ground state.  
Thus, they correspond to two different ways of rotating the symmetry broken ground state, either in the $YZ$ plane or the $XZ$ plane, matching the two different Goldstone modes above.

Therefore, for a square region, which has $4$ such corners, we expect a correction
of
\be
\approx -0.0496 \ln(\ell).
\ee
As we saw above, numerical results disagree with this, suggesting some nontrivial effects that are not accounted for by this framework.

\section{Summary and Discussion}

In this paper, we have presented several computational methods to calculate
the entanglement properties of lattice statistical models, in
dimensionality greater than one, in a systematic
manner. The stochastic series expansion QMC and high temperature expansions
are finite-temperature methods. Since the former works with finite systems,
it is possible to calculate the ground state properties by going to 
sufficiently low temperatures. The latter is a series expansion, defined in the
thermodynamic limit. It can, in principle, be extrapolated to $T \to 0$ limit
if there is no finite temperature phase transition. An interesting finding of 
our paper is that the limits $T\to 0$ and $L\to\infty$ need not commute
in these calculations.

We have also developed computational methods that work directly at $T=0$.
The valence bond QMC is an exact ground state projection method for a finite
system. The Ising series expansions represent an expansion in exchange
anisotropy around a given classical state. Using these methods we have
calculated the area law associated with the boundary, as well as subleading
logarithmic terms associated with the corners and the bulk. Whenever
quantities are calculated by two different methods, there is good
quantitative agreement.

We have also discussed a mean-field calculation of the entanglement properties
as well as the results expected from non-interacting Bosons. Here, we find
a surprise that the numerical results disagree with simple expectations.
The mean-field state, where all spins on one sublattice are equally entangled
with spins on the other sublattice, predicts a bulk log term of $2\ln{\ell}$,
whereas our valence bond QMC results give $c\ln{\ell}$, with $c= 0.74\pm0.02$.
The latter is closer to a recent spin-wave calculation, where an estimate
of $c=0.92$ was obtained.\cite{yalegroup} The discrepancy is even more puzzling for the
corner terms. If long-wavelength spin-waves act as non-interacting Bosons,
they should contribute a log term with a coefficient of $-0.0496$. 
That number should be compared with $-0.080\pm 0.008$ obtained in Ising series 
expansions and $\approx -0.1$ in the QMC studies. These disagreements
suggest the need for further theoretical study.

\section*{Acknowledgements}
We acknowledge collaboration related to the development of the VB QMC loop/ratio algorithm by H.~G.~Evertz.  
We are indebted to I.~Gonzalez for providing DMRG results for Fig.~\ref{mconv}.
Support was provided by NSF grant number  DMR-1004231 and NSERC of Canada.
The simulations were made possible by the facilities of the Shared Hierarchical Academic Research Computing Network (SHARCNET:www.sharcnet.ca) and Compute/Calcul Canada.

\end{document}